# Bridging the Quantum Divide: Aligning Academic and Industry Goals in Software Engineering

Jake Zappin
*William & Mary*
Williamsburg, Virginia, USA
azappin@wm.edu

Trevor Stalnaker
*William & Mary*
Williamsburg, Virginia, USA
twstalnaker@wm.edu

Oscar Chaparro
*William & Mary*
Williamsburg, Virginia, USA
oscarch@wm.edu

Denys Poshyvanyk
*William & Mary*
Williamsburg, Virginia, USA
denys@cs.wm.edu

*Abstract*—This position paper examines the substantial divide between academia and industry within quantum software engineering. For example, while academic research related to debugging and testing predominantly focuses on a limited subset of primarily quantum-specific issues, industry practitioners face a broader range of practical concerns, including software integration, compatibility, and real-world implementation hurdles. This disconnect mainly arises due to academia's limited access to industry practices and the often confidential, competitive nature of quantum development in commercial settings. As a result, academic advancements often fail to translate into actionable tools and methodologies that meet industry needs. By analyzing discussions within quantum developer forums, we identify key gaps in focus and resource availability that hinder progress on both sides. We propose collaborative efforts aimed at developing practical tools, methodologies, and best practices to bridge this divide, enabling academia to address the application-driven needs of industry and fostering a more aligned, sustainable ecosystem for quantum software development.

## I. INTRODUCTION

Software development for quantum computers is still nascent [1–3]. Unlike classical computing, quantum computing (QC) lacks well-established paradigms, tools, and practices that support robust software engineering (SE). Quantum software developers currently face several unique challenges that complicate the process of building, testing, and debugging applications. Quantum mechanics introduces probabilistic behaviors and complex entanglements, which are non-intuitive for traditional developers [3, 4]. The steep learning curve, combined with other factors such as the lack of mature tools and frameworks, very few consistently maintained open-source software (OSS) projects, and a tight-lipped development community due to commercial and strategic interests, means that quantum software development is often fragmented and insular, limiting opportunities for shared learning and collaborative advancements [3, 5–8].

The importance of structured SE in QC cannot be overstated. As quantum hardware becomes more sophisticated, the complexity of quantum software will increase accordingly. Early investment in rigorous SE practices is essential to build a foundation for reliable, scalable, and high-quality quantum applications. By establishing foundational best practices and robust toolsets now, we can reduce the barriers to future quantum software development, improve the quality of applications, and accelerate the adoption of quantum technologies [9–11].

Academic researchers have recently begun to explore QC from a SE perspective. However, much of this work has been theoretical, without direct observation of the practices, challenges, or methodologies employed by quantum developers in industry (*e.g.*, private companies and government entities working on QC and software development). For instance, Akbar *et al.* [7] propose a readiness model to help organizations assess their preparedness for transitioning from classical to quantum software development, highlighting critical areas for capability building. Moguel *et al.* [8] drew lessons from the classical software crisis in the 1960s and 1970s, proposing a roadmap for quantum SE (QSE) that emphasizes structured practices to avoid similar pitfalls. Zhao *et al.* [1] provided a survey that defines QSE as a distinct field, emphasizing the need for robust engineering principles borrowed from classical SE to support the lifecycle of quantum applications.

While academic research is laying a critical foundation for QSE, there is a noticeable disconnect between these theoretical frameworks and the realities faced by quantum developers in industry. Unlike classical software development, where academic findings translate more readily into practical tools and methodologies [12, 13], QSE has seen limited uptake of academic solutions within industry settings as evidenced by our research [14] where we observed virtually no usage by practitioners of quantum testing and debugging tools developed in academia. Moreover, in classical software development, partnerships between industry and academia are relatively commonplace [15–17]. This is not the case in the world of quantum development, with the exception of a handful of partnerships between start-ups and academic consortia. Factors such as the secrecy surrounding quantum research (often having national security implications), the siloed nature of quantum start-ups, the lack of large, well-maintained open-source quantum software projects, and the strategic competition among major players have all contributed to the divide between academic research and industry practices.

We believe that advancing QC and software development requires a stronger partnership between academia and industry. By fostering collaboration and sharing insights from both sectors, we can accelerate the development of essential tools, methodologies, and best practices needed to address the unique challenges of QSE. In this position paper, we first provide an overview of the current industry landscape in quantum software and computing, followed by our observations from our research that highlight some aspects of the divide between academia and industry in SE practices. Next, we explain why we believe that it is important for academia and industry to work more closely. Finally, while acknowledging the significant obstacles to overcome, we propose potential first steps and solutions to bridge this divide.

## II. THE CURRENT LANDSCAPE

### A. Industry Landscape

The QC industry is currently led by IBM, which dominates both market share and influence in quantum software and hardware development [?]. IBM's Qiskit platform has established itself as the leading framework for quantum programming, providing a relatively robust toolkit that is widely used across academic and industrial settings [18, 19]. This dominance is further bolstered by IBM's partnerships with a handful of academic institutions through the IBM Quantum Initiative program, although these collaborations have primarily focused on advancing hardware capabilities rather than software development and engineering [20].

While other companies like Google [21], Microsoft [22], and Amazon [23] have developed competing quantum platforms, IBM's Qiskit remains dominant in adoption, features, and documentation [18, 19, 21–24]. In addition to these technology giants, a number of quantum-focused companies are working to advance QC and software development. Xanadu, for instance, has made strides in photonic quantum computing and offers the PennyLane library, which focuses on hybrid quantum-classical (HQC) machine learning [25, 26]. Rigetti [27] and IonQ [28] are pioneering quantum hardware platforms, each with a proprietary quantum cloud service aimed at facilitating quantum application development. Companies like Atom Computing [29], PsiQuantum [30], and Quantum Brilliance [31] are working on specialized quantum software solutions and technologies, such as trapped-ion and photonic QC, while Quantinuum [32] and IQM [33] are developing software solutions to enhance quantum algorithm performance. Despite these advances, most of these companies operate as startups with relatively small user bases, limiting their impact on the broader quantum computing ecosystem, at least for now.

### B. Challenges Facing Academia and Industry Collaboration

The QC industry faces unique challenges that hinder collaboration and slow software innovation. The industry is a highly secretive, competitive environment motivated by being the first-to-market with a quantum breakthrough. Many companies require its developers to sign non-disclosure agreements, preventing open discussions of best practices and other SE issues. This secrecy is compounded by a small developer community—which we estimate to be only a few thousand professionals globally—who appear to work in isolation on specialized projects [34], limiting opportunities for community-driven progress like OSS projects contributions.

A major driver of this secrecy is national security. Much of the quantum development in industry, especially in the U.S., is funded and guided by government contracts with large technology companies like IBM and Google, start-ups such as PsiQuantum and IonQ, and even traditional defense contractors like Boeing and Lockheed Martin participating [35–37]. This national security focus restricts collaboration and isolates industry efforts from academia. Indeed, researchers in academia have had apparent difficulty in getting industry quantum developers to participate in their research projects and data gathering efforts [38, 39].

For academia, an industry environment cloaked in secrecy drastically limits insight into the practices and challenges of actual practitioners. Researchers are unable to discern the real-world challenges in quantum software development faced by industry developers. This lack of visibility into industry has deepened the divide between academic advancements and industry needs, hindering academia's progress in developing effective quantum SE research.

## III. OBSERVATIONS OF THE QUANTUM DIVIDE

Our research into QSE [14] reveals significant evidence of a divide between academia and industry within the quantum field. The insights we gathered expose gaps in resources, focus, and approaches that likely hinder both academic and industry progress in quantum software development. These observations are detailed below.

### A. Scarcity of Quality OSS Repositories for HQC Applications

Our analysis of quantum developer forums revealed a notable scarcity of robust, well-maintained OSS repositories for hybrid quantum- classical (HQC) applications [14]. Among the repositories surveyed, only 540 were explicitly labeled as hybrid applications, with over 50% limited to research artifacts, educational exercises, or simple reproductions of example code from documentation. Alarmingly, 91% of these repositories showed minimal activity, with fewer than 15 commits or 10 reported issues, and most had not been updated within the past year of our study.

Furthermore, only six repositories utilized formal issue tracking, and documentation quality was often poor, with commit messages rarely describing the bugs being addressed. This lack of comprehensive, publicly accessible, and actively maintained OSS projects creates a significant barrier for academic researchers, who lack access to real-world quantum development practices for study. Consequently, academia struggles to observe, document, and analyze prevailing industry practices, limiting the ability to generate insights into the practical challenges faced by developers in the quantum field. As a result, industry practitioners often face unaddressed difficulties due to the absence of accessible, real-world examples that could otherwise serve as invaluable resources for learning and problem-solving.

### B. Neglect of Hybrid System Bugs in Academia vs. Industry's Focus on Hybrid Applications

Academic research in QSE has primarily focused on purely quantum issues in quantum development, with minimal attention to the unique bugs and integration challenges that arise specifically in HQC systems [2, 40–44]. This limited focus fails to address the complexities developers face when combining quantum and classical components, where hybrid-specific bugs frequently emerge. Our analysis of 531 real-world developer issues found that HQC integration challenges accounted for 91 reported issues (17.1%), highlighting a key pain point that remains largely unexplored in academic studies [14, 45, 46]. These issues often involve data encoding mismatches, synchronization errors, and compatibility issues unique to hybrid architectures, which can significantly hinder the reliability and scalability of HQC applications [14].

Moreover, industry has placed significant emphasis on developing functional HQC applications, particularly in fields like machine learning, where hybrid models appear to offer tangible benefits [47, 48]. Consequently, developers frequently encounter hybrid-specific failures, including platform interoperability issues, faulty parameter encoding, and optimization errors, which collectively accounted for a substantial portion of our dataset [14]. Notably, these types of cross-domain issues—where errors arise at the intersection of quantum and classical components—comprised 17.1% of all reported issues in our study, further underscoring the need for targeted research in hybrid quantum-classical computing [14]. This disparity highlights the disconnect in focus between efforts of academia and industry's real-world challenges, leaving critical hybrid quantum software engineering bugs and issues largely unaddressed by research efforts in academia.

*C. Disparities in Focus on Error Types and Bug Patterns*

Our investigation into discussions within quantum developer forums revealed additional differences between the types of issues encountered by industry practitioners and those emphasized in academic research. Practitioners frequently face classical issues while developing quantum applications, such as integration challenges, compatibility issues, and platform-specific limitations. These issues collectively accounted for 207 reported issues (39.0%) in our dataset [14]. Developers often express frustration with the complexities of integrating quantum algorithms with classical systems, managing compatibility between quantum libraries, and adapting code for different hardware environments. For instance, library and platform issues (116 instances) were among the most frequently cited problems, with developers struggling to resolve version mismatches, API changes, and inconsistent support across quantum computing frameworks. These challenges are further exacerbated by limited interoperability and inconsistent functionalities across available quantum tools.

In contrast, academic research has primarily focused on idealized, isolated quantum bugs, such as qubit ordering and entanglement errors [2, 40–44]. While these studies contribute to fundamental advancements in QC, they fail to address the broader engineering challenges faced by practitioners, particularly in HQC environments where classical and quantum software components are combined to exploit quantum advantages while maintaining classical computational efficiency. Our findings suggest that academia has largely overlooked issues related to integration, dependency handling, and compatibility testing, among others, that practitioners routinely encounter. Without greater collaboration between academia and industry, QSE research risks remaining disconnected from the pressing needs of practitioners, slowing progress toward scalable and reliable quantum applications.

*D. Broad Spectrum of Issue Categories in Developer Forums*

Quantum developer forums highlight diverse challenges, particularly in debugging, testing, and integrating quantum-classical components. This spectrum includes software faults, configuration issues, library and platform limitations, hardware/simulator constraints, and developer errors [14]. The diversity of issues faced in practice extends beyond the narrow focus of most academic research, which often emphasizes algorithm optimization, circuit design, and quantum-specific error correction [2, 40–44]. This focus on theoretical advancements leaves many practical challenges, such as debugging and testing in hybrid systems, unaddressed. This gap underscores the need for academia to broaden its scope to encompass more practical issues faced by quantum software developers in real-world environments.

*E. Perspectives on Platform vs. Application Development*

Our findings reveal a notable disconnect between academic research objectives and the immediate, practical needs of quantum application developers. Academic research has primarily approached quantum software issues from the perspective of quantum platform developers, focusing on identifying and addressing quantum-specific bugs in frameworks like Qiskit and Cirq [2, 40–44]. While application developers do face challenges related to these platforms, academic research often overlooks the real-world needs of developers focused on creating user-oriented solutions rather than enhancing underlying quantum infrastructure [14]. Platform developers are primarily concerned with the reliability and capabilities of QC frameworks, whereas application developers are focused on using these platforms to solve domain-specific problems. This divergence highlights a critical research gap, as the nuanced, practical challenges of application developers—key to advancing real-world quantum applications—remains largely unaddressed by academia.

## IV. THE IMPORTANCE OF COLLABORATION BETWEEN ACADEMIC AND INDUSTRY

As QC advances, bridging the gap between academic research and industry practice is essential to overcome the unique challenges and unlock the full potential of this emerging field. Both academia and industry bring distinct strengths to QSE. Academia excels in theoretical exploration, long-term research, and training future developers. It also has the time and capacity to develop the tools and methodologies that industry developers lack. Meanwhile, industry has the resources, practical focus, and direct access to quantum hardware to advance applied QC. It is also the first to encounter challenges and obstacles that are not easily overcome without dedicated research. Increasing the collaboration between these sectors is vital to creating a sustainable and efficient ecosystem that drives meaningful progress in QC.

- *Advancing Quantum Software Engineering Practices:* Collaboration between academia and industry can shape quantum software engineering practices suited to unique challenges like probabilistic states, resource limits, and hybrid systems. Without industry input, academic work risks being overly theoretical. Joint efforts can create standards and tools that are both practical and rigorously tested in real scenarios.
- *Accelerating Tool and Framework Development:* The lack of specialized tools for testing and debugging quantum software is a significant barrier. While academics often create limited-scope tools, industry needs solutions that integrate with existing workflows. Collaboration can bridge this gap, producing robust tools that reduce debugging time and improve reliability.
- *Addressing Knowledge Gaps and Training:* The rapid pace of quantum computing requires developers skilled in both theory and practical application. Academia provides foundational knowledge, while industry offers exposure to the latest technologies. Joint efforts can equip developers with real-world skills, bridging the talent gap in quantum.
- *Fostering Innovation and Solving Practical Problems:* Collaboration combines academic creativity with industry pragmatism, driving innovation. Academia's research complements industry's focus on immediate challenges, which could lead to breakthroughs that neither sector could achieve alone.

## V. Bridging the Quantum Divide

To bridge the academia-industry divide in quantum software engineering (QSE), we propose several initiatives to encourage collaboration and shared progress.

### A. Confidential Research Collaboratives

Establishing confidential research collaboratives between universities and companies can help balance the need for confidentiality with the benefits of collaboration. These initiatives would involve academic researchers and industry professionals working together on short- and long-term projects under structured non-disclosure agreements to protect proprietary information. Through these collaboratives, academic researchers would gain controlled access to real-world industry challenges, enabling them to tailor their research to address practical issues relevant to quantum software development. Such frameworks would allow information sharing while respecting confidentiality, enabling academics to contribute meaningfully to industry-relevant solutions in areas such as testing, debugging, and integration.

### B. Government-Funded Industry-Academia Partnerships

Given the national security interest in quantum technology, government-funded partnerships can be an effective way to bridge the academic-industry divide. These partnerships, modeled after public-private initiatives in other high-tech sectors, would provide secure funding for foundational QSE needs, such as error mitigation, testing frameworks, and hybrid system compatibility. By overseeing these partnerships, government agencies could establish secure, legal frameworks for sharing sensitive information, allowing academia and industry to collaborate on high-impact projects without compromising security. Such partnerships would enable academics to gain insight into industry challenges, directly informing the development of practical tools and methodologies.

### C. Establishing QSE Consortiums Within Universities

Academic institutions should consider establishing on-campus consortiums focused on QSE, targeting practical issues like error correction, hybrid integration, and testing frameworks. Industry partners would contribute funding and guidance, ensuring research aligns with real-world needs. Unlike confidential research collaboratives, consortiums would foster a broader, semi-open environment, encouraging shared learning and resource development among multiple academic and industry partners. While many quantum consortiums already exist, they primarily focus on quantum physics rather than SE. These consortiums could train students in QSE, preparing them to meet industry demands, and serve as incubators for start-ups and new ideas. Events such as joint seminars, project showcases, and conferences would further promote knowledge sharing and build a collaborative community.

### D. Collaboration on Quality OSS Quantum Projects

To bridge academic research with industry needs, a focused effort on OSS quantum projects is essential. Collaborative OSS projects could provide a shared space for academia, industry, and other contributors to work on quantum software solutions for real-world challenges. Building repositories for practical quantum applications—such as error correction libraries, HQC frameworks, or specialized algorithm toolkits—would create resources that benefit both sectors. OSS projects would enhance academia's understanding of industry practices, helping researchers create relevant tools and methodologies, while providing training opportunities for upcoming quantum developers. We propose a blend of industry, academic, and government funding to support these initiatives. Government funding, in particular, would encourage broad contributions, help maintain neutrality, and sustain projects over the long term, supporting regular updates, documentation, and community support.

## VI. Conclusion

A clear disconnect exists between academia's theoretical focus in QC and the practical, application-driven needs of industry. Bridging this gap through stronger collaboration and a shift toward real-world challenges in research is essential to advance QSE. By aligning research efforts with industry demands, we can accelerate the development of impactful tools and methodologies that support the growth of QC applications.

## REFERENCES

[1] J. Zhao, "Quantum software engineering: Landscapes and horizons," *arXiv preprint arXiv:2007.07047*, 2020.

[2] M. Paltenghi and M. Pradel, "Bugs in quantum computing platforms: An empirical study," *Proceedings of the ACM on Programming Languages*, vol. 6, no. OOPSLA1, pp. 1–27, 2022.

[3] S. Resch and U. R. Karpuzcu, "Quantum computing: an overview across the system stack," *arXiv preprint arXiv:1905.07240*, 2019.

[4] P. Zhao, J. Zhao, and L. Ma, "Identifying bug patterns in quantum programs," in *2021 IEEE/ACM 2nd International Workshop on Quantum Software Engineering (Q-SE)*, 2021, pp. 16–21.

[5] J. Preskill, "Quantum computing in the nisq era and beyond," *Quantum*, vol. 2, p. 79, 2018.

[6] T. Lubinski, S. Johri, P. Varosy, J. Coleman, L. Zhao, J. Necaise, C. H. Baldwin, K. Mayer, and T. Proctor, "Application-oriented performance benchmarks for quantum computing," *IEEE Transactions on Quantum Engineering*, vol. 4, pp. 1–32, 2023.

[7] M. A. Akbar, S. Rafi, and A. A. Khan, "Classical to quantum software migration journey begins: a conceptual readiness model," in *International Conference on Product-Focused Software Process Improvement*, 2022, pp. 563–573.

[8] E. Moguel, J. Berrocal, J. García-Alonso, and J. M. Murillo, "A roadmap for quantum software engineering: Applying the lessons learned from the classics." in *Q-SET@ QCE*, 2020, pp. 5–13.

[9] J. M. Murillo, J. Garcia-Alonso, E. Moguel, J. Barzen, F. Leymann, S. Ali, T. Yue, P. Arcaini, R. P. Castillo, I. G. R. de Guzmán *et al.*, "Challenges of quantum software engineering for the next decade: The road ahead," *arXiv preprint arXiv:2404.06825*, 2024.

[10] M. De Stefano, F. Pecorelli, D. Di Nucci, F. Palomba, and A. De Lucia, "Software engineering for quantum programming: How far are we?" *Journal of Systems and Software*, vol. 190, p. 111326, 2022.

[11] C. Oyeniran, A. O. Adewusi, A. G. Adeleke, L. A. Akwawa, and C. F. Azubuko, "Advancements in quantum computing and their implications for software development," *Computer Science & IT Research Journal*, vol. 4, no. 3, pp. 577–593, 2023.

[12] D. I. Sjøberg, J. E. Hannay, O. Hansen, V. B. Kampenes, A. Karahasanovic, N.-K. Liborg, and A. C. Rekdal, "A survey of controlled experiments in software engineering," *IEEE transactions on software engineering*, vol. 31, no. 9, pp. 733–753, 2005.

[13] B. Fitzgerald and K.-J. Stol, "Continuous software engineering and beyond: trends and challenges," in *Proceedings of the 1st International Workshop on rapid continuous software engineering*, 2014, pp. 1–9.

[14] J. Zappin, T. Stalnaker, O. Chaparro, and D. Poshyvanyk, "When quantum meets classical: Characterizinghybrid quantum-classical issues discussed in developer forums," in *Proceedings of the 47th IEEE/ACM International Conference on Software Engineering (ICSE'25)*, p. to appear.

[15] T. Gorschek, P. Garre, S. Larsson, and C. Wohlin, "A model for technology transfer in practice," *IEEE software*, vol. 23, no. 6, pp. 88–95, 2006.

[16] J. Feller, P. Finnegan, and O. Nilsson, "Open innovation and public administration: transformational typologies and business model impacts," *European Journal of Information Systems*, vol. 20, no. 3, pp. 358–374, 2011.

[17] J. Noll, S. Beecham, and I. Richardson, "Global software development and collaboration: barriers and solutions," *ACM inroads*, vol. 1, no. 3, pp. 66–78, 2011.

[18] A. Javadi-Abhari, M. Treinish, K. Krsulich, C. J. Wood, J. Lishman, J. Gacon, S. Martiel, P. D. Nation, L. S. Bishop, A. W. Cross *et al.*, "Quantum computing with qiskit," *arXiv preprint arXiv:2405.08810*, 2024.

[19] G. Aleksandrowicz, T. Alexander, P. Barkoutsos, L. Bello, Y. Ben-Haim, D. Bucher, F. J. Cabrera-Hernández, J. Carballo-Franquis, A. Chen, C.-F. Chen *et al.*, "Qiskit: An open-source framework for quantum computing," *Accessed on: Mar*, vol. 16, 2019.

[20] "Ibm quantum initiative," 11 2024. [Online]. Available: https://quantum.ncsu.edu/

[21] G. A. Q. Team, "Cirq," 2018. [Online]. Available: https://github.com/quantumlib/Cirq

[22] "Introduction to the quantum programming language q#," 10 2024. [Online]. Available: https://learn.microsoft.com/en-us/azure/quantum/qsharp-overview

[23] "Amazon braket," 11 2024. [Online]. Available: https://aws.amazon.com/braket/

[24] "cuquantum sdk: A high-performance library for accelerating quantum science," 11 2024. [Online]. Available: https://docs.nvidia.com/cuda/cuquantum/latest/index.html

[25] X. AI, "Xanadu, linkedin profile," 2023. [Online]. Available: https://ca.linkedin.com/company/xanaduai

[26] ——, "Templates - pennylane 0.33.0 documentation," 2023. [Online]. Available: https://docs.pennylane.ai/en/stable/introduction/templates.html

[27] "Rigetti - our approach," 11 2024. [Online]. Available: https://www.rigetti.com/about-rigetti-computing

[28] "Ionq - our mission," 11 2024. [Online]. Available: https://ionq.com/company

[29] "Atom computing," 11 2024. [Online]. Available: https://atom-computing.com/

[30] "Psiquantum," 11 2024. [Online]. Available: https://www.psiquantum.com/

[31] "Quantum brilliance," 11 2024. [Online]. Available: https://quantumbrilliance.com/

[32] "Quantinuum," 11 2024. [Online]. Available: https://www.quantinuum.com/

[33] "Iqm," 11 2024. [Online]. Available: https://www.meetiqm.com/

[34] K. Upadhyay, V. Chhetri, A. Siddique, and U. Farooq, "Analyzing the evolution and maintenance of quantum computing repositories," *arXiv preprint arXiv:2501.06894*, 2025.

[35] "Darpa advances psiquantum to second phase of utility-scale quantum computing program," 6 2024. [Online]. Available: https://www.psiquantum.com/news-import/darpa-advances-psiquantum-to-second-phase-of-utility-scale-quantum-computing-program

[36] "Ibm intends to partner with fermilab's sqms center to advance critical quantum information science initiatives," 7 2024. [Online]. Available: https://newsroom.ibm.com/2024-07-18-IBM-Intends-to-Partner-with-Fermilabs-SQMS-Center-to-Advance-Critical-Quantum-Information-Science-Initiatives

[37] "Ionq announces largest 2024 u.s. quantum contract award," 9 2024. [Online]. Available: https://ionq.com/news/ionq-announces-largest-2024-u-s-quantum-contract-award-of-usd54-5m-with

[38] L. Jimnez-Navajas, F. Bhler, F. Leymann, M. Piattini, and D. Vietz, "Quantum software development: A survey," *Quantum Information and Computation*, vol. 24, no. 7&8, pp. 0609–0642, 2024.

[39] M. De Stefano, F. Pecorelli, F. Palomba, D. Taibi, D. Di Nucci, and A. De Lucia, "Quantum software engineering issues and challenges: Insights from practitioners," in *Quantum Software: Aspects of Theory and System Design*, 2024, pp. 337–355.

[40] H. Li, F. Khomh, L. Tidjon *et al.*, "Bug characteristics in quantum software ecosystem," *arXiv preprint arXiv:2204.11965*, 2022.

[41] J. Luo, P. Zhao, Z. Miao, S. Lan, and J. Zhao, "A comprehensive study of bug fixes in quantum programs," in *2022 IEEE International Conference on Software Analysis, Evolution and Reengineering (SANER)*. IEEE, 2022, pp. 1239–1246.

[42] P. Zhao, X. Wu, J. Luo, Z. Li, and J. Zhao, "An empirical study of bugs in quantum machine learning frameworks," *arXiv preprint arXiv:2306.06369*, 2023.

[43] P. Zhao, J. Zhao, Z. Miao, and S. Lan, "Bugs4q: A benchmark of real bugs for quantum programs," in *2021 36th IEEE/ACM International Conference on Automated Software Engineering (ASE)*, 2021, pp. 1373–1376.

[44] P. K. Nayak, K. V. Kher, M. B. Chandra, M. Rao, and L. Zhang, "Q-pac: Automated detection of quantum bug-fix patterns," *arXiv preprint arXiv:2311.17705*, 2023.

[45] S. Endo, Z. Cai, S. C. Benjamin, and X. Yuan, "Hybrid quantum-classical algorithms and quantum error mitigation," *Journal of the Physical Society of Japan*, vol. 90, no. 3, 2021.

[46] F. Phillipson, N. Neumann, and R. Wezeman, "Classification of hybrid quantum-classical computing," in *International Conference on Computational Science*, 2023, pp. 18–33.

[47] G. De Luca, "A survey of nisq era hybrid quantum-classical machine learning research," *Journal of Artificial Intelligence and Technology*, vol. 2, no. 1, pp. 9–15, 2022.

[48] L. Wossnig, "Quantum machine learning for classical data," *arXiv preprint arXiv:2105.03684*, 2021.